\documentclass{article}
\usepackage[pdftex]{color,graphicx}
\usepackage{hyperref}
\usepackage{html}
\begin{document}
\author{Kirana Kumara P\\Centre for Product Design and Manufacturing,\\Indian Institute of Science,\\Bangalore, Karnataka 560012, India\\\textit{Email: kiranakumarap@gmail.com}
}
\title{Demonstrating the Usefulness of CAELinux for Computer Aided Engineering using an Example of the Three Dimensional Reconstruction of a Pig Liver}
\date{}
\maketitle
\begin{abstract}
CAELinux is a Linux distribution which is bundled with free software packages related to Computer Aided Engineering (CAE). The free software packages include software that can build a three dimensional solid model, programs that can mesh a geometry, software for carrying out Finite Element Analysis (FEA), programs that can carry out image processing etc. Present work has two goals: 1) To give a brief description of CAELinux 2) To demonstrate that CAELinux could be useful for Computer Aided Engineering, using an example of the three dimensional reconstruction of a pig liver from a stack of CT-scan  images. One can note that instead of using CAELinux, using commercial software for reconstructing the liver would cost a lot of money. One can also note that CAELinux is a free and open source operating system and all software packages that are included in the operating system are also free. Hence one can conclude that CAELinux could be a very useful tool in application areas like surgical simulation which require three dimensional reconstructions of biological organs. Also, one can see that CAELinux could be a very useful tool for Computer Aided Engineering, in general.
\end{abstract}
\section{Introduction}
Most of the popular Computer Aided Engineering (CAE) software packages (e.g., SolidWorks, Altair HyperMesh, ANSYS, MSC Nastran, ANSYS FLUENT, COMSOL Multiphysics, Amira, 3D-DOCTOR) are expensive. But there exist a few good free (and/or open source) software packages that are quite comparable with commercial software. For example, the free software package SALOME can do 3D modelling, meshing and pre/post-processing, the software Code$\textunderscore$Aster can do Finite Element Analysis (FEA), OpenFOAM can do Computational Fluid Dynamics (CFD) analysis, Elmer can perform multiphysical simulations, ITK-SNAP can do segmentation of medical images. It is cumbersome to separately download all these free software packages. Also, some of the versions (or releases) of some of the above software packages may not be compatible with some of the versions of some of the operating systems. Also, one has to ensure that all the required libraries are available and all dependencies are resolved. Also, software packages have to be individually configured. Needless to say, if one chooses to individually compile each of these open source software packages (from source) for a given operating system and a given hardware, it would take a lot of time and effort.

Hence it is better if one can get these free CAE software packages at one place, all bundled together. It is even better if the operating system itself is free and open source, so that these CAE software packages and the operating system can be bundled together to produce a seamless system. This would avoid version and library dependencies/conflicts. In this case, the free CAE software packages become the inbuilt `packages' of the free operating system. CAELinux~\cite{1} is an operating system that is created to achieve these objectives.

CAELinux is built on top of the popular Linux distribution Ubuntu. CAELinux is free and open source software (operating system). It comes with numerous CAE software packages, all free. In addition to CAE software packages, it has many software packages that are useful for engineers and scientists, e.g., Maxima, which is a computer algebra system; Scilab, Octave and FreeMat, which are free alternatives for the popular commercial software MATLAB; Gnuplot, which is a graphing utility; LaTeX, LyX, TeXworks and Texmaker, which are useful for typesetting. It also has inbuilt development tools like C, C++, Fortran and Python compilers. Since it is built on top of Ubuntu, it includes all software packages that come bundled with Ubuntu, e.g., OpenOffice.org. Additionally, one can seamlessly download any software package that is available in the huge Ubuntu repository. CAELinux can also be run live from a DVD, without the necessity of installing it into the hard disk. If desired, one can install it into a hard disk; CAELinux DVD includes tools which can automatically analyze and partition the chosen hard disk drive and install CAELinux side by side with other operating systems such as Windows; tools can automatically create/update boot loader also. CAELinux is also available through the virtual machine hosting service Amazon Elastic Cloud Computing (EC2)~\cite{2}; but as of now, CAELinux EC2 image is released as beta version.

Software packages that are bundled with CAELinux are useful for a wide variety of CAE applications. But, as a demonstration, the present work concerns itself only on the problem of reconstructing a biological organ from an image stack that was obtained through CT-scan. In this work, a three dimensional surface model of a porcine liver is extracted from an image stack that was freely available on the Internet. Only CAELinux, with no other additional software, is used for the purpose. The successful reconstruction of the liver demonstrates that CAELinux can be used for extracting three dimensional biological organs from image stacks obtained through CT-scan.

Present work aims to achieve two goals: 1) To give a brief description of CAELinux (this goal has already been achieved by the description of CAELinux given in this section) 2) To demonstrate that CAELinux could be useful for Computer Aided Engineering, using an example of the three dimensional reconstruction of a pig liver from a stack of CT-scan images (this is to be demonstrated from the next section onwards).

Essentially, present work demonstrates how useful the free and open source software CAELinux could be, by taking just one topic (in CAE) of the three dimensional reconstruction which needs just a few of the many software packages available within CAELinux. One can see that following the same lines as followed in the present paper, one could demonstrate that the numerous software packages available within CAELinux could be useful for a wide variety of purposes (which may not have been even mentioned in this paper also). Hence one can see that it would be possible to demonstrate that CAELinux could be a single free alternative to various expensive commercial software packages that are used in many of the different areas under Computer Aided Engineering. The brief description of CAELinux given in this section, and the subsequent demonstration that the CAELinux can successfully be used to reconstruct the pig liver (to be demonstrated from the next section onwards), substantiate this idea.
 
\section{Related Works}
One can see that the free and open source software CAELinux, by its very nature, could be very useful in many different areas related to CAE. A short review of CAELinux may be found in \cite{3}.

Present author has not come across any source in the literature which uses CAELinux to reconstruct surface models of three dimensional biological organs from image sequences that are obtained through CT-scan. However, the practice of using free software packages to reconstruct biological organs like liver and kidney may be found in the author’s previous works \cite{4,5,6}. Present paper and author’s previous works \cite{4,5,6} both use the same software packages, i.e., ImageJ~\cite{7,8,9}, ITK-SNAP~\cite{10,11}, MeshLab~\cite{12,13}, to perform the three dimensional reconstruction. But the present work is different from these previous works in that the previous works do not use CAELinux while the present work uses CAELinux alone; CAELinux contains ImageJ, ITK-SNAP and MeshLab.

Of course, \cite{4,5,6} are not the only works found in the literature that use free software packages to extract the geometry of biological organs. For example, \cite{14} deals with the three dimensional reconstruction of liver slice images using the free software called MITK~\cite{15}. Also, one can find many works that deal with the three dimensional reconstruction of biological organs using commercial software packages, e.g., \cite{16,17,18,19}.
\section{Three Dimensional Reconstruction of a Pig Liver}
In this section, a pig liver is reconstructed (extracted) from an image stack. The image stack was downloaded from \cite{20}. The image stack was downloaded sometime back when it was available in the website. But now, the website does not contain the image stack. The image stack consists of 147 two dimensional images obtained through CT-scan. CT-scan images were obtained for one half of the pig liver only; other half of the liver may be considered to be symmetrical to the first half. One can refer to \cite{21} for further details on the image stack.

The procedure used here to reconstruct the 3D model of the liver from the image stack is explained in detail in \cite{4}.

\begin{figure}
\begin{center}
\includegraphics[width=0.5\textwidth]{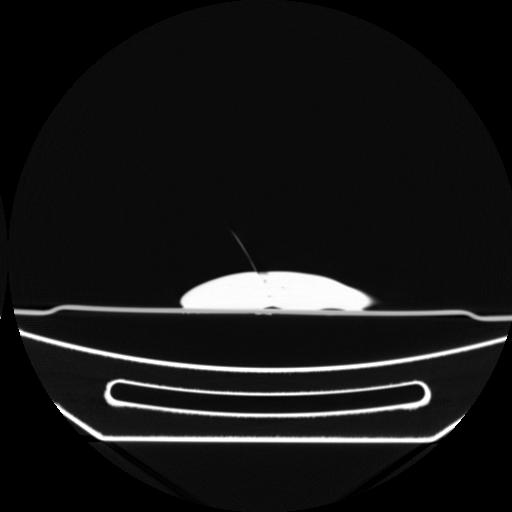}
\caption{The 25th Image (Before Image Arithmetic)}          
\label{figure1}
\end{center} 
\end{figure}

\begin{figure}
\begin{center}
\includegraphics[width=0.5\textwidth]{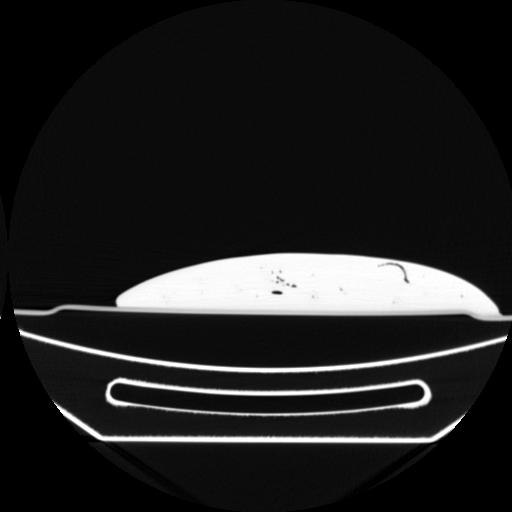}
\caption{The 75th Image (Before Image Arithmetic)}          
\label{figure2}
\end{center} 
\end{figure}

First, the image stack is imported into ImageJ, an image processing software available within CAELinux. Figure~\ref{figure1} and Figure~\ref{figure2} show the 25th and the 75th image of the image stack respectively.

\begin{figure}
\begin{center}
\includegraphics[width=0.5\textwidth]{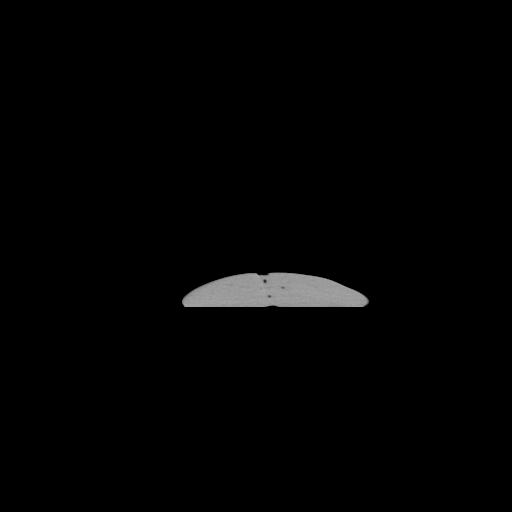}
\caption{The 25th Image (After Image Arithmetic)}          
\label{figure3}
\end{center} 
\end{figure}

\begin{figure}
\begin{center}
\includegraphics[width=0.5\textwidth]{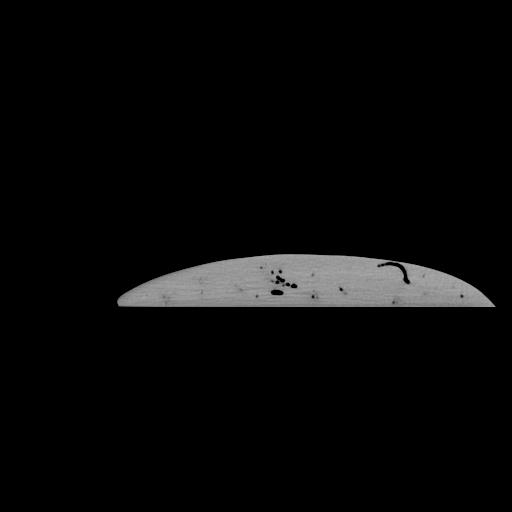}
\caption{The 75th Image (After Image Arithmetic)}          
\label{figure4}
\end{center} 
\end{figure}

In Figure~\ref{figure1} and Figure~\ref{figure2}, the liver is located around the centre of the images and the bottom portion of the images shows the part of the experimental set up supporting the liver. If one considers the region where the liver is located, one can see that intensity of the pixels belonging to the liver is different from the intensity of the pixels that do not belong to the liver; the pixels belonging to the liver are brighter. ImageJ can do image arithmetic; hence one can consider just the area that is close to the liver, and one can continue subtracting a fixed intensity value from each of the pixels in this area, for each of the images in the image stack, until only those pixels that belong to the liver have positive intensity values. Figure~\ref{figure3} and Figure~\ref{figure4} show the 25th and the 75th image of the image stack, after conducting image arithmetic.

Now, the image stack is mirrored, and both the original and the mirrored images are retained in the stack. This is achieved by `stack reverser' and `concatenate' features available in ImageJ. Now the image stack contains 294 images (instead of 147 images) and it represents the full liver (not half of the liver). The processed image stack is now saved.

Now, the processed image stack is imported into ITK-SNAP, a software available within CAELinux. ITK-SNAP is used to do segmentation of each of the images in the image stack. Finally, the segmented volume is exported as a surface mesh composed of triangles.

\begin{figure}
\begin{center}
\includegraphics[width=0.5\textwidth]{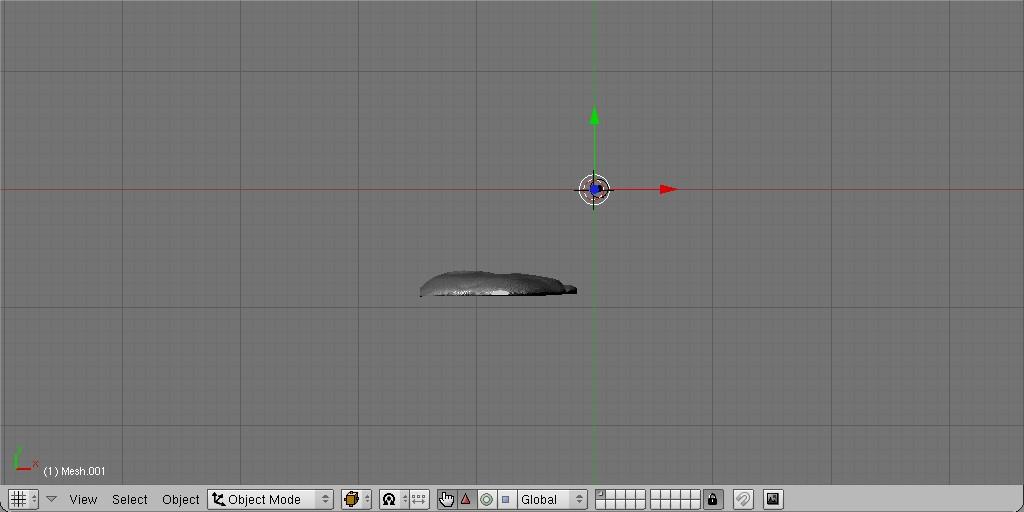}
\caption{The 3D Model of the Liver (274553 Faces)}          
\label{figure5}
\end{center} 
\end{figure}

The surface mesh obtained is the three dimensional model of the porcine liver. The model, as seen in Blender~\cite{22} (yet another software available within CAELinux), is shown in Figure~\ref{figure5}.

\begin{figure}
\begin{center}
\includegraphics[width=0.5\textwidth]{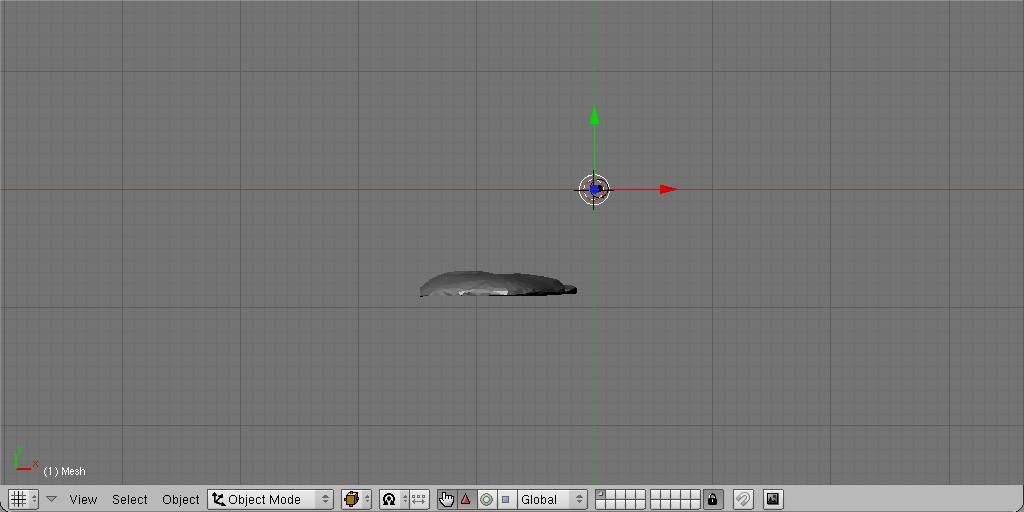}
\caption{The 3D Model of the Liver (1500 Faces)}          
\label{figure6}
\end{center} 
\end{figure}

Usually, a surface mesh obtained this way is made up of too many surface triangles, e.g., the surface mesh shown in Figure~\ref{figure5} is made up of 274553 triangular faces. Too many number of faces is a problem if one wants to use the three dimensional model in a finite element analysis. CAELinux includes software known as MeshLab which can be of help here. MeshLab can approximate a 3D model represented by a large number of triangular faces to a 3D model that is represented by a smaller number of faces. Here, MeshLab is used to reduce the total number of surface triangles from 274553 to 1500. Now, the model of the liver represented by 1500 faces, as seen in Blender, is shown in Figure~\ref{figure6}. Comparing Figure~\ref{figure6} with Figure~\ref{figure5}, one can see that the liver can satisfactorily be described by just 1500 faces; hence it is not necessary that one should have too many numbers of faces to accurately describe a complicated geometry.

Whole of the above procedure may also be followed to freely obtain patient specific 3D models of biological organs from patient specific scanned images. Also, here, it is worthwhile to note that ``Visible Human Data Set''~\cite{23} is a full collection of transverse CT, MR and cryosection images of representative male and female cadavers. The whole data set may be downloaded, after obtaining a licence. One need not pay any fee to get the licence. Hence, using CAELinux and this data set, and by following the procedure illustrated in the present section, one can reconstruct (extract) a representative three dimensional geometry of any human organ for free. An example of the reconstruction of human kidney from the Visible Human Data Set may be found in \cite{6}.

As regards to validation of the reconstructed model of the liver, the reconstructed model is not validated here. The software packages used in this work to reconstruct the liver are quite established and it is reasonable to assume that they do not produce erroneous results. Although it is possible to perform the same reconstruction (as the one carried out in the present work) using some well established commercial software and compare the liver thus reconstructed with the one obtained here to establish the validity of the procedure followed here, this step may not be required since the present work just uses a few software packages all of which would have undergone testing prior to their release.
\section{Concluding Remarks}
Previous section, through the live example of reconstructing a liver, demonstrates that CAELinux alone (without the need of any extra software packages) can reconstruct (i.e., extract) three dimensional geometry of biological organs from the two dimensional images that are obtained through CT-scan. The task involves image arithmetic, segmentation, and exporting the segmented volume as a surface mesh. CAELinux includes software packages that are necessary to carry out all these subtasks. Since CAELinux is free and open source software (operating system), one does not need to pay for commercial software packages that can do the reconstruction task. It may be noted that commercial software packages which can accomplish the same task are not inexpensive. Application areas like surgery planning and surgery simulation require three dimensional reconstructions of biological organs and hence, CAELinux could be a great alternative to commercial software packages which can accomplish this task.

Three dimensional reconstructions is only one of the numerous topics in image processing, and image processing is just one of the innumerable topics in Computer Aided Engineering. Present paper concerns itself only about three dimensional reconstructions which is just one of the innumerable areas in Computer Aided Engineering where CAELinux could be of great use. Also, present paper utilizes extremely small number of the total available CAE software packages that are available within the CAELinux distribution. However, present work demonstrates how useful the free and open source software CAELinux could be, by taking just one topic (in CAE) of the three dimensional reconstruction which needs just a few of the many software packages available within CAELinux. Also, following the same lines as followed in the present paper, one could demonstrate that the numerous software packages (which may not have been even mentioned in this paper) available within CAELinux could be useful for a variety of purposes (which are not covered in the present paper); hence one could demonstrate that CAELinux could be a single free alternative to various expensive commercial software packages that are used in many of the different areas that fall under Computer Aided Engineering; the brief description of CAELinux given in the first section of this paper, and the subsequent demonstration that the CAELinux (and the CAELinux alone) can successfully be used to reconstruct the pig liver, substantiate this idea.

Future work is to use CAELinux to solve diverse scientific and engineering problems. This would help to understand the usefulness of CAELinux with the subsequent use of this free and open source software package (or operating system) to economically solve complex scientific and engineering problems.
\section*{Acknowledgments}
Author is grateful to the Robotics Lab, Department of Mechanical Engineering and Centre for Product Design and Manufacturing, Indian Institute of Science, Bangalore, INDIA, for providing the necessary infrastructure to carry out this work.

Author thanks Prof. Ashitava Ghosal, Department of Mechanical Engineering and Centre for Product Design and Manufacturing, Indian Institute of Science, Bangalore, INDIA, for giving some initial ideas.

\end{document}